\documentclass[sigconf]{acmart}

\usepackage{amsmath}
\usepackage{graphicx}

\setcopyright{acmlicensed}
\copyrightyear{2026}
\acmYear{2026}
\setcopyright{cc}
\setcctype{by}
\acmConference[CHI EA '26]{Extended Abstracts of the 2026 CHI Conference on Human Factors in Computing Systems}{April 13--17, 2026}{Barcelona, Spain}
\acmBooktitle{Extended Abstracts of the 2026 CHI Conference on Human Factors in Computing Systems (CHI EA '26), April 13--17, 2026, Barcelona, Spain}
\acmDOI{10.1145/3772363.3798758}
\acmISBN{979-8-4007-2281-3/2026/04}

\begin{document}

\title{One Kiss: Emojis as Agents of Genre Flux in Generative Comics}

\author{Xiruo Wang}
\authornote{First author and corresponding author.}
\orcid{0009-0005-1631-3073}
\email{xiruo.wang.22@ucl.ac.uk}
\affiliation{%
  \institution{University College London}
    \city{London}
    \state{}
    \country{United Kingdom}
}

\author{Xinyi Jiang}
\orcid{0009-0004-0698-0000}
\email{xinyijiang2002@163.com}
\affiliation{%
  \institution{Tsinghua University}
  \city{Beijing}
  \country{China}
}

\author{Ziqi Lyu}
\orcid{0009-0003-8826-9708}
\email{LzqLeisure@gmail.com}
\affiliation{%
  \institution{Beijing Forestry University}
  \city{Beijing}
  \country{China}
}

\begin{abstract}
Generative AI has made visual storytelling widely accessible, yet current prompt-based interactions often force users into a trade-off between precise control and creative flow. We present \textbf{One Kiss}, a co-creative comic generation system that introduces ``Affective Steering''. Instead of writing text prompts, users guide the tone of their story through emoji inputs, whose semantic ambiguity becomes a resource rather than a limitation. Unlike traditional text-to-image tools that rely on explicit descriptions, One Kiss uses a dual-stream input in which users define structural
pacing by sketching panel frames and set atmospheric tone by
pairing keywords with emojis. This mechanism enables ``Genre Flux,'' where emotional inputs accumulate across panels and gradually shift the genre of a story. A preliminary study ($N=6$) suggests that this soft steering approach may reframe the user's role from prompt engineer to narrative director, with ambiguity serving as a source of creative surprise rather than a loss of control.
\end{abstract}

\ccsdesc[500]{Human-centered computing~HCI design and evaluation methods}
\ccsdesc[500]{Human-centered computing~Interactive systems and tools}
\ccsdesc[300]{Human-centered computing~Interaction design}

\keywords{generative AI, co-creation, emoji, affective computing, 
comic generation, human-AI interaction}

\begin{teaserfigure}
  \centering
  \includegraphics[width=\textwidth]{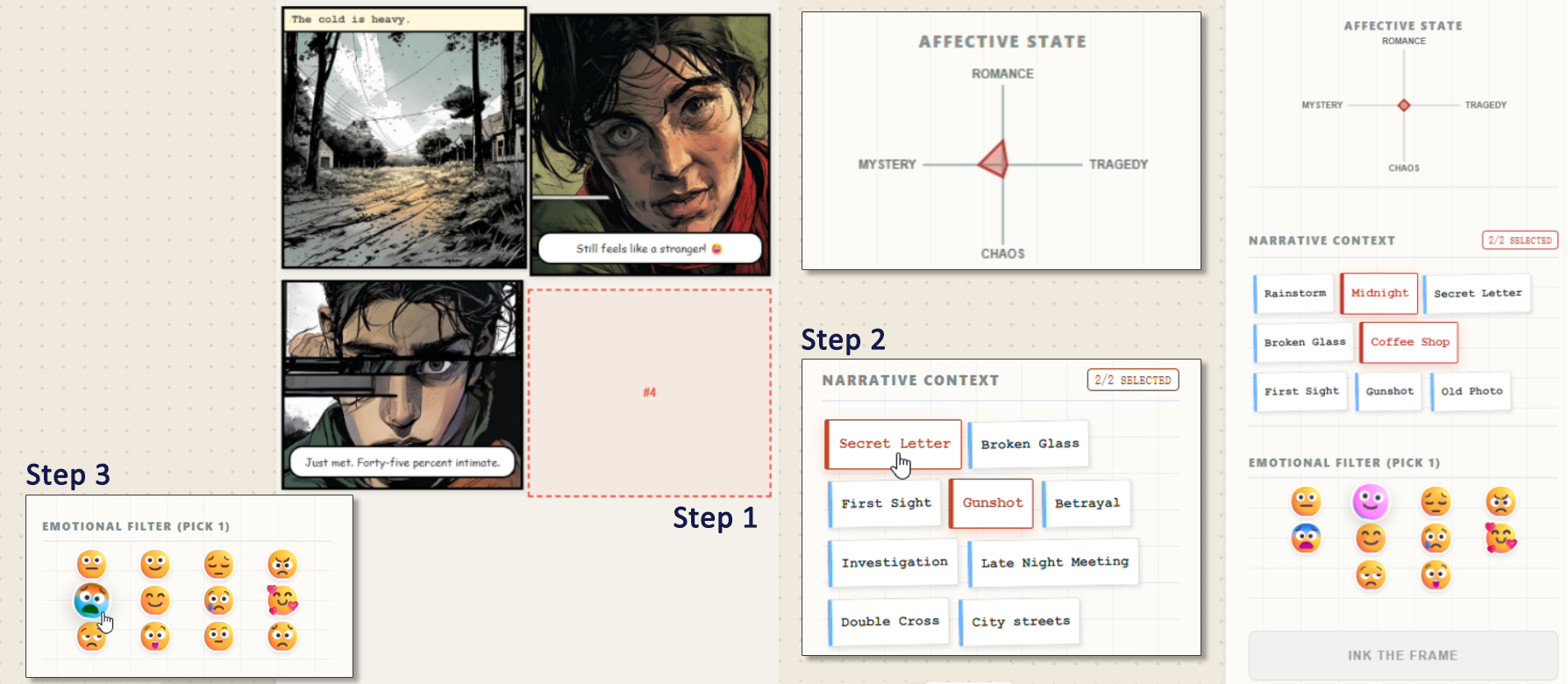}
  \caption{The One Kiss Interface. Users define structure via \textbf{Spatial Pacing} (sketching the frame for Panel 4) and modulate tone via \textbf{Affective Injection} (pairing keywords like ``Gunshot'' with a ``Worried'' emoji). The system's \textbf{Genre Flux} algorithm updates the affective radar chart (top right), shifting the narrative style toward ``Mystery'' and ``Chaos'' for the next generation.}
  \Description{A screenshot of the One Kiss interface. The center shows a canvas with comic panels. The right sidebar shows controls for selecting keywords and emojis. A radar chart at the top right visualizes the current emotional state.}
  \label{fig:teaser}
\end{teaserfigure}

\maketitle

\section{Introduction}

Text-to-image models now allow novices to produce complex imagery without traditional artistic skills \cite{chung2022talebrush}. However, the prevailing interaction paradigm of text-based prompting imposes a substantial cognitive load on users. To maintain narrative consistency or achieve a specific atmospheric tone, users are often forced to engage in laborious prompt engineering, an iterative process that resembles debugging code rather than creative expression \cite{zamfirescu2023johnny}. This gap widens when users attempt to articulate implicit affective qualities, such as ``a melancholic silence,'' rather than explicit physical descriptions. While text is precise for specifying objects, it is often too rigid for the non-verbal qualities of emotional storytelling.

We take a different approach: rather than increasing prompt precision, we treat ambiguity as a design resource. Drawing on Gaver et al.'s framework of ambiguity as a resource for design \cite{gaver2003ambiguity}, we suggest that semantic ambiguity can open spaces for interpretation and surprise. In digital communication, emojis carry rich emotional meaning in a compact, non-verbal form~\cite{alshenqeeti2016emojis}. Unlike text prompts that specify what the AI should draw, emojis offer a way to steer how the narrative feels, providing a form of control that feels intuitive rather than technical.

In this work, we introduce \textbf{One Kiss}, an interactive comic
generation system built around what we call \textbf{Affective
Steering}, designed for creative exploration rather than production.
The system decouples narrative control into two streams. First, users
define structural rhythm through spatial sketching. Second, they
shape narrative tone through emoji inputs. This mechanism enables
\textbf{Genre Flux}, where affective inputs accumulate and gradually
shift the genre of the story. Our preliminary evaluation ($N=6$)
suggests that this approach reduces the anxiety of blank-page
generation and may reframe the user's role from prompt engineer to
narrative director. This paper contributes the One Kiss prototype
and empirical insights into how soft control maintains creative flow
during AI co-creation.

\section{Related Work}

\subsection{Steering Interaction in Co-Creative Narrative Systems}

Text prompting remains the dominant interface for generative AI,
yet non-expert users often struggle to express affective intentions
through text alone~\cite{zamfirescu2023johnny}. Co-creative systems
have responded by introducing alternative forms of steering, but
these have focused on two areas: narrative structure and visual
composition.

On the structural side, \textit{TaleBrush} lets users sketch
fortune curves to control how a character's arc rises and
falls across a story~\cite{chung2022talebrush}, while dialogue-based
systems like \textit{ID.8}~\cite{antony2024id8} and
\textit{``1001 Nights''}~\cite{Fu2025like} use planning
conversations and role-play agents to support plot development. On
the visual side, \textit{Opal}~\cite{liu2022opal} and \textit{AI
Instruments}~\cite{riche2025aiinstruments} offer multimodal cues
such as layouts and reference images to guide image generation
without relying on text alone.

What these approaches share is that they steer \emph{what happens} in a
story or \emph{what a scene looks like}, but not \emph{how the
story feels over time}. Even in sequential visual story generation, where models like StoryDALL-E~\cite{maharana2022storydalle} maintain cross-frame
visual consistency, each panel is generated in a tonally independent
way. There is no mechanism for emotional signals to accumulate
across a sequence and gradually reshape the genre. One Kiss
addresses this gap.

\subsection{Ambiguity as a Resource for Affective Steering}

To steer tone rather than structure, we need an input modality that
is expressive but deliberately imprecise. Gaver et al.'s framing
of ambiguity as a design resource~\cite{gaver2003ambiguity}
suggests that interpretive openness can invite engagement rather
than frustrate it. Emojis fit this description: large-scale sentiment analyses have shown that emojis carry quantifiable emotional valence \cite{kraljnovak2015sentiment}, and experimental evidence
confirms that they effectively convey emotionality in digital
contexts \cite{erle2021emojis}, yet their meaning remains
sufficiently open to support interpretive
flexibility \cite{alshenqeeti2016emojis}.

One Kiss applies this through Affective Steering: users pair
keywords with emojis, and these pairings accumulate over time,
shifting the visual genre rather than specifying plot events.

\begin{figure*}[t]
  \centering
  \includegraphics[width=\textwidth]{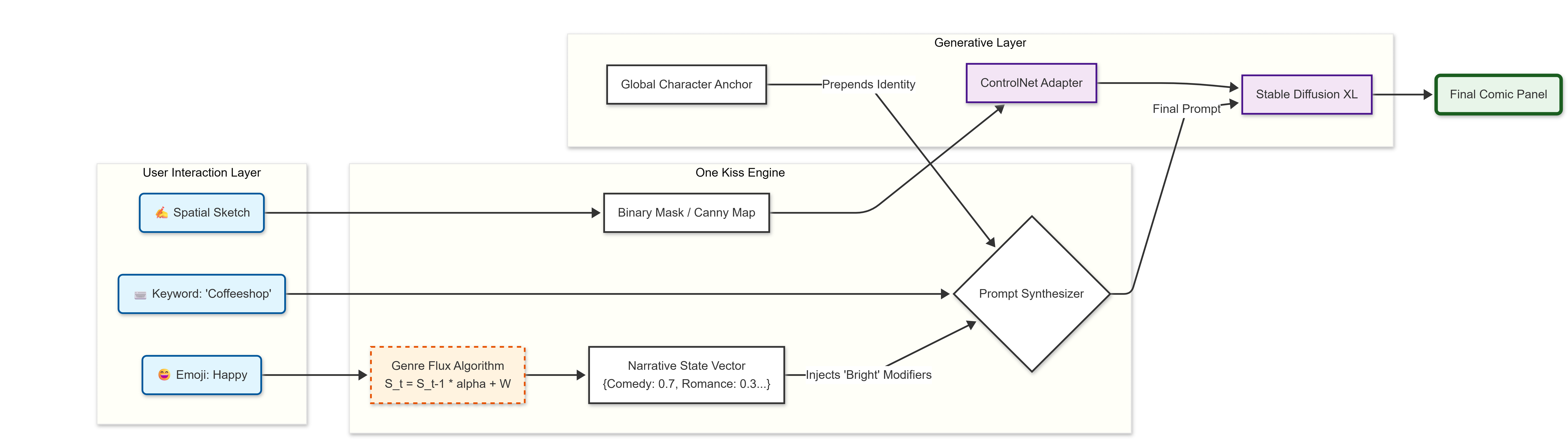} 
  \caption{Technical Workflow. The system processes inputs through two parallel streams: (1) A \textbf{Spatial Stream} that converts sketches into ControlNet guidance, and (2) an \textbf{Affective Stream} where the Genre Flux Algorithm converts emojis into dynamic style modifiers. These streams converge in the Prompt Synthesizer to guide the Stable Diffusion model.}
  \Description{A flowchart diagram illustrating the system architecture.}
  \label{fig:pipeline}
\end{figure*}

\section{System Implementation}

We developed One Kiss as an interactive research prototype to
implement Affective Steering. The system architecture consists of a
dual-stream frontend for multimodal input and a state-aware backend
for narrative accumulation.

\subsection{The Dual-Stream Interface}

\textbf{Spatial Pacing Stream:} Users begin a session by sketching
a bounding box and selecting an initial narrative keyword to
generate the first panel. Subsequent panels are added iteratively,
with each sketch defining the visual rhythm for that beat. The
dimensions of the box serve as a structural prompt, instructing the
image generator to adjust composition complexity. For example, a
wide, short box biases the model towards panoramic, cinematic
shots, whereas a tall, narrow box triggers close-up character
portraits. This allows users to control the camera language without
technical vocabulary.

\textbf{Affective Injection Stream:} Rather than writing text
prompts for every detail, users pair a keyword from the interface
vocabulary (e.g., ``Gunshot'', ``Broken Glass'') with an emoji
(e.g., ``Wilted Flower''). This pairing drives the generation. The
interface provides visual feedback through a particle system that
changes color based on the injected emoji, so users can see how
their input affects the system's state.

\subsection{Algorithmic Steering and Genre Flux}

Unlike stateless generation models that treat each prompt in
isolation, One Kiss maintains a dynamic \textbf{Narrative State
Vector ($V_t$)}. Central to this is the Genre Flux Algorithm, which
tracks the cumulative emotional trajectory of the story.

Let $S_t$ be the emotional score vector at panel $t$, comprising
four dimensions: $\{Romance, Tragedy, Chaos, Mystery\}$. These four dimensions were chosen for the current prototype to cover a broad range of genre directions; they are not intended as a fixed taxonomy and could be extended in future iterations. The state update rule is
defined as:

\begin{equation}
    S_{t} = S_{t-1} \cdot \alpha + (W_{kw} + W_{emoji}) \cdot \beta
\end{equation}

Where $\alpha$ is a decay factor empirically tuned during pilot
testing to balance narrative continuity and responsiveness to new
inputs; a value of 0.8 was found to allow gradual genre drift
without jarring tonal shifts. $\beta$ is a rarity multiplier
ranging from 1.0 to 3.0, derived from the semantic rarity of the
user's keyword choice, on the assumption that unusual inputs signal
intentional narrative pivots.

The system continuously monitors $S_t$. When a specific dimension
dominates (e.g., $S_{tragedy} > \theta$, where $\theta$ is the flux
threshold), the system activates a \textbf{Style Modifier}. This
modifier does not act merely as a color filter. It injects
genre-specific constraints into both the prompt and negative prompt,
influencing compositional density and lighting contrast. For
instance, crossing the Tragedy threshold might inject ``monochrome
blue palette, high contrast shadows, film noir grain'' while
simultaneously suppressing ``bright colors, cheerful expressions,''
making the genre shift visible.

To address the common challenge of character hallucination in
sequential generation, the system implements a \textbf{Global
Character Anchor}. A fixed visual description prompt is prepended
to every generation request, ensuring that while the genre style
shifts dynamically, the protagonist's identity features remain
consistent across panels. The current prototype focuses on
single-protagonist narratives, allowing the Genre Flux mechanism
to maintain a coherent visual identity throughout the comic. The
protagonist's visual description is specified by the user at the
start of each session.

\section{Preliminary Evaluation}

We conducted an exploratory user study to test Affective Steering
in practice and understand how users perceive the
control-ambiguity trade-off.

\subsection{Participants and Procedure}

We recruited 6 participants (3 female, 3 male, aged 22--26) with
varying levels of experience in generative AI tools. The study was
conducted remotely using a think-aloud protocol. The procedure
consisted of three phases. First, a 10-minute tutorial introduced
users to the spatial sketching and emoji interface. Second, users
were given a creative task: create a 6-panel comic that starts with
a neutral tone and ``drifts'' into a specific genre (e.g., Horror,
Romance, or Tragedy) using only the provided tools. Finally, a
post-task semi-structured interview explored their sense of agency
and creative flow.

\subsection{Results: From Engineering to Directing}

Participants used the emoji inputs to steer narrative trajectories.
Two authors independently reviewed the 36 generated panels,
assessing whether the dominant visual style, as indicated by color palette, lighting contrast, and compositional density, had shifted
to match each participant's stated target genre by the fourth panel.
This analysis confirmed that all six participants successfully
triggered a Genre Flux, with disagreements resolved through
discussion.

Qualitative feedback pointed to a shift in how users thought about
the interaction. One participant (P2) noted that they did not have
to explicitly describe the sadness or instruct the AI to make the
scene dark. Instead, dropping a wilted flower emoji caused the
lighting to shift naturally. They described this interaction as akin
to ``giving an actor a mood cue rather than writing a technical
script.'' This suggests that semantic ambiguity lets users work at
the level of narrative mood rather than struggling with word choice
and prompt syntax. Four of six participants described the
interaction as feeling more like directing a story than prompting a
system, and several noted that compared to their prior experience
with text-to-image tools, the emoji-based input felt less
intimidating when starting from a blank canvas.

\subsection{Serendipity and Perceived Agency}

Participants did not see the ambiguity of emojis as a drawback.
Instead, they reported that the lack of granular control encouraged
a sense of discovery. For example, P4 paired a ``Clown'' emoji with
a ``Rain'' keyword, expecting a literal interpretation. The system
instead generated a surreal and chaotic background with distorted
colors, which the user embraced as a creative plot twist. At the
sequence level, P3 used emoji inputs to steer a six-panel comic
through a thriller arc that concluded with the protagonists
kissing, an outcome they had not planned but described as surprising
yet coherent. This suggests that Genre Flux can produce surprises
that span the whole story, not just individual panels.

These observations suggest a shift in how agency is distributed.
Users give up control over pixel-level rendering while keeping
influence over pacing and genre. Several participants also described
a sense of agency over the narrative direction. This may relate to
the fact that the primary structural decisions, such as panel layout
and emotional pivots through emoji choice, originated from the
user, even though the pixel-level rendering was automated.

Participants also identified tensions in the spatial pacing stream.
One user noted that early keyword choices constrained later scenes.
Selecting ``Coffee Shop'' as an initial context caused subsequent
panels to remain anchored to that setting, even when the intended
narrative had moved on. Another observed that consecutive panels
often defaulted to similar close-up compositions, and wanted more
control over how camera framing varied across panels. These
observations point to a limitation of the current prototype. While
the affective stream supports gradual drift, the spatial stream
does not yet support the same kind of gradual evolution. Given the
small sample size of six participants, these findings should be
interpreted as exploratory.

\section{Discussion}

Beyond the specific prototype, One Kiss raises a broader question:
how can semantic ambiguity serve as a resource in generative
systems?

\subsection{The Tension Between Precision and Ambiguity}

This work highlights a trade-off between lexical precision and
affective expression. Traditional prompt-based systems prioritize
object-level accuracy (e.g., ``a red chair''), which often fails to
capture atmospheric nuance without extensive engineering. In
contrast, Affective Steering sacrifices this granular control for
high-level narrative resonance. Our study suggests that for creative
exploration, users gravitated toward the ``soft control'' of emojis,
which act as semantic scaffolds. This suggests a shift from an
``architectural'' view of creativity toward something closer to
gardening, planting seeds and seeing what grows.

\subsection{Balancing Consistency and Drift}

A critical tension in our design is the balance between narrative
drift and character consistency. While the Genre Flux algorithm
intentionally destabilizes the visual style, the Global Character
Anchor is essential to prevent the narrative from collapsing into
incoherence. Future iterations will need to refine this balance,
perhaps by allowing users to lock specific visual elements that
should remain immune to the flux.

\section{Conclusion and Future Work}

We introduced One Kiss, a system that uses emoji ambiguity to steer
the genre of generative comics. Our preliminary results suggest
that this soft steering approach can support creative flow and
unexpected discovery. Future work will explore fine-grained steering
mechanisms, such as continuous intensity sliders for emoji weights,
and investigate therapeutic applications where non-verbal emoji
inputs could help users express difficult emotions through digital
storytelling.

\bibliographystyle{ACM-Reference-Format}
\bibliography{sample-base} 

\end{document}